\documentclass[smallcondensed,natbib]{svjour3}    

\smartqed  

\usepackage{graphicx}
\usepackage{ulem}             
\usepackage{amssymb,amsmath,amsfonts}
\usepackage{graphicx}
\usepackage{color}

\newcommand{\NN}{{\mathbb N}}

\newcommand{\lam}{{\lambda}}
\newcommand{\Lam}{{\Lambda}}
\newcommand{\hzeta}{{\hat\zeta}}

\newcommand{\bH}{{\overline H}}

\newcommand{\bu}{{\overline u}}
\newcommand{\bz}{{\overline z}}
\newcommand{\bZ}{{\overline Z}}
\newcommand{\bw}{{\overline w}}
\newcommand{\bW}{{\overline W}}
\newcommand{\tH}{{\widetilde H}}
\newcommand{\smu}{{\sqrt\mu}}
\newcommand{\cP}{{\cal P}}
\newcommand{\cO}{{\cal O}}

\newcommand{\be}{\begin{equation}}
\newcommand{\ee}{\end{equation}}

\newcommand{\der}[2]{\frac{d#1}{d#2}}
\newcommand{\dron}[2]{\frac{\partial#1}{\partial#2}}

\begin{document}
\title{Influence of the coorbital resonance on the  rotation of the Trojan satellites of Saturn}


\author{Philippe Robutel \and  Nicolas Rambaux \and  Maryame El Moutamid}


\institute{P. Robutel \and N. Rambaux \and M. El Moutamid \at
              ASD, IMCCE-CNRS UMR8028, Observatoire de Paris, Paris, France \\
               \email{robutel@imcce.fr} 
               \and N. Rambaux \at 
               Universit\'e Pierre et Marie Curie - UPMC, France
               \and M. El Moutamid
               \at LESIA, Observatoire de Paris, Paris, France 
               }     

\maketitle
\date{\today}

\begin{abstract}
The Cassini spacecraft collects high resolution images of the saturnian satellites and reveals the surface of these new worlds. The shape and rotation of the satellites can be determined from the Cassini Imaging Science Subsystem data, employing limb coordinates and stereogrammetric control points. This is the case for Epimetheus \citep{TiThBu2009} that opens elaboration of  new rotational models \citep{TiThBu2009, No2010, RoRaCa2011}. Especially, Epimetheus is characterized by its horseshoe shape orbit and the presence of the swap has to be introduce explicitly into rotational models.
During its journey in the saturnian system, Cassini spacecraft accumulates the observational data of the other satellites and it will be possible to determine the rotational parameters of several of them. To prepare these future observations, we built rotational models of the coorbital  (also called Trojan) satellites Telesto, Calypso, Helene, and Polydeuces, in addition to Janus and Epimetheus. Indeed, Telesto and Calypso orbit around the $L_4$ and $L_5$ Lagrange points of Saturn-Tethys while Helene and Polydeuces are coorbital of Dione. The goal of this study is to understand how the departure from the Keplerian motion induced by the perturbations of the coorbital body, influences  the rotation of these satellites. To this aim, we introduce explicitly the perturbation in the rotational equations by using the formalism developed by \citep{Ed1977} to represent the coorbital motions, and so we describe the rotational motion of the coorbitals, Janus and Epimetheus included, in compact form.

\keywords{Rotation \and Coorbitals \and Libration \and Secondary resonance \and Saturn satellites}

\end{abstract}

\maketitle


\section{Introduction}
\label{sec:intro}

The space mission Cassini orbiting Saturn since 2004 provides a lot of data and notably images of the surface of over 20 satellites \citep{Thomas2010}. From these images methods based on limb profile provide the satellites shape \citep{Thomas-etal1998, ThoBuHeSqVePoTu1997, Thomas2010}. To combine the different images taken at different time, it is crucial to assume a rotational motion of the satellites. Thus, the rotational determination is obtained as a by-product result.
Usually for almost satellites a synchronous uniform rotational motion seems to be enough to fit the images \citep{ThoBuHeSqVePoTu1997}. However, for specific cases such as Janus and Epimetheus more elaborated rotational models are required \citep{TiThBu2009} because of the swap in their orbital motion.
This swap is due to the 1:1 mean motion resonance where every 4 years these satellites approach and swap their orbits by a few tens of kilometers. The resulting orbits in the rotating reference frame are an horseshoe shape orbits.

In the Saturnian system, four additional coorbital satellites (i.e. in 1:1 orbital resonance) are currently known. They are Calypso and Telesto that are coorbital of Tethys, and, Helene and Polydeuces, coorbital of Dione. Contrary to Janus and Epimetheus, their orbits describe a tadpole shape in the rotating reference frame (Fig.~\ref{fig:orb}). Their masses are very small with respect to their main satellites Tethys and Dione, whereas Janus and Epimetheus have a comparable size.
According to \textit{Horizons} ephemeris \citep{Gioetal1996}, the mass ratio between Janus and  Epimetheus  is about $3.604$, while the masses of Telesto, Calypso, and Helene are several tens of thousand times smaller than the one of Tethys and Dione.  We note that Polydeuces' mass  is currently unknown, but it is expected much smaller than the one of Helene because the mean radius of Polydeuces is less than $2$ kilometers \citep{PoThoWeRi2007}. Consequently, the orbital motion of these four small  satellites may be deduced from the restricted three body problem including Saturn and one of these small bodies, Tethys or Dione. In this framework, the small satellites are located close to the $L_4$ and $L_5$ Lagrange points of the main satellites, executing a tadpole orbit around one of the Lagrange points (see \cite{ChNaMo} and references therein).
The objective of the present paper is to investigate the influence of the orbital oscillations resulting from the 1:1 orbital resonance, on the rotational motion of these satellites by developing a general method based on the Hamiltonian approach and parametrization of the orbital motion.

The coorbital satellites have irregular shape that probably results from their original accretion or their impact history \citep{PoThoWeRi2007, Thomas2010}. The shape can be approximated by an ellipsoid obtained from the best fit to the numerical shape model developed by Thomas (2010). But, for the rotation, the main parameters of interest are the moments of inertia. They can be deduced from the  semi-axes of the ellipsoid by assuming an homogenous interior. The error can be estimated of the order of 30 \% for Janus (see Section  \ref{sec:discu}), so we could explore a large range of values around the homogenous ellipsoidal shape. In addition, we assume that the rotation of these satellites is synchronous as in \cite{PoThoWeRi2007} and \cite{Thomas2010}  because this is the most expected state due to tidal dissipation (e.g. \citeauthor{Peale99}, \citeyear{Peale99}).

In \cite{RoRaCa2011} a perturbative approach has been used to solve the equations of the rotational motion by assuming that the orbital trajectories of Janus and Epimetheus were quasiperiodic. This method allows to determine and to measure the influence of the coorbital resonance on  the rotation of theses satellites. Here, we propose to apply such a method to compute efficiently the rotational motion of these satellites. As the satellites present a 1:1 orbital resonance and 1:1 spin-orbit resonance, first we clarify vocabulary used in this paper. 
On one hand, the 1:1 mean motion resonance generates tadpole or horseshoe orbits along which the difference of the longitude of the two satellites are perturbed (called usually libration) with an orbital frequency $\nu$ that we called the coorbital frequency.
On the other hand, the 1:1 spin-orbit resonance also causes a libration corresponding to an oscillation of the rotation angle of the satellite with respect to an uniform rotation. For quasiperiodic motion, the frequency that depends on the amplitude of the oscillation is called proper libration frequency. This frequency will be denoted $\sigma$ at the center of libration (see e.g. \cite{Wisdom2004}). The term libration is used only to describe the quasiperiodic rotational motion of the satellites.

The paper is divided as follows: first we describe the representation of the orbital motion that we use to model the coorbital satellites. We then present a Hamiltonian formulation of the rotational problem and discuss on non-resonant and secondary resonant cases. These secondary resonant cases result in commensurability between the libration frequency and orbital frequency. They appear for some values of the triaxiality of the satellites and we detail the 1:1 and 1:2 secondary resonances. In particular, we show how the overlap of resonances associated with the coorbital frequency generates relatively large chaotic regions. Finally, we discuss our results in prevision of future observations and the assumption made in this paper.

\section{Approximation of the orbital motion}
\label{sec:orbit}
 \medskip
In this paper, we restrict the orbital model to the circular restricted three-body problem. It is within this simplified framework that the phenomena we want to highlight will appear the most clearly. 
Let us consider two bodies of masses $m_0$ (Saturn) and $m_1$ (Dione or Tethys) with  $m_0\gg m_1$ and  
\be 
\mu = \frac{m_1}{m_0+m_1}.
\label{eq:mu}
\ee
The body of mass $m_1$ describes a circular orbit of radius $r_0$ centered on the most massive  body with an angular velocity (mean motion) equal to $n$. The origin of the reference frame is located at the massive body (the planet), and  polar coordinates are assigned to the two other bodies: $(r_0,v_p)$ to the main satellite, and  $(r,v)$ to the trojan.

\cite{Ed1977}, in his paper dedicated to the dynamics of the Jovian Trojans, derives asymptotical solutions of the trojan motion in the case of the elliptic restricted three body problem expending the series up to order two in $\smu$.  
Limiting \'Erdi's theory to the first order in $\smu$ and assuming circular revolutions for the two massive bodies, the coordinates $(r,v)$ can be approximated by these expressions\footnote{See the appendix for the relationship of these expressions with the \citeauthor{Ed1977}'s theory (\citeyear{Ed1977}).}:

\begin{eqnarray}
 r(t)\phantom{xx}   &= &\phantom{-}r_0\left[
  1 +\sqrt{\mu}\
    \rho\cos(v_p(t) + \zeta(t) + \psi) -\frac{2}{3n} \der{\zeta(t)}{t}
                              \right]   \label{eq:sol_orb_r},\\
v(t) \phantom{xx}  &= &\phantom{-}v_p(t) + \zeta(t) -2\sqrt{\mu} \rho\sin( v_p(t) + \zeta(t) + \psi)  \label{eq:sol_orb_v} ,\\
 v_p(t) \phantom{x,}&= &\phantom{-} n t + v_{p,0}   \label{eq:sol_orb_vp} , \\
 \der{^2\zeta}{\, t^2}(t)  & = & - 3\mu n^2\left(
               1 - (2-2\cos \zeta(t))^{-3/2}\right)\sin \zeta (t)   \label{eq:sol_orb_zeta}.
     \end{eqnarray}
The coefficients $(\rho,\psi,v_{p,0})$ are constant parameters linked with the initial conditions. 
This model  provides a quasiperiodic approximation of the motion of the massless body in the coorbital 1:1 mean motion resonance, which is valid for a tadpole orbit as well as for a horseshoe obit. 

The previous expressions split  in two components: a fast component whose amplitude is controlled by $\rho$, and a slow component governed by the function $\zeta$.   This function represents the temporal variations of $v-v_p$ averaged on one orbital period $2\pi/n$. An usual approximation of this quantity is given by the solution of the differential equation (\ref{eq:sol_orb_zeta}) (see \cite{Kevorkian1970}, \cite{SaYo1988}, and also \cite{Morais2001} for an averaged Hamiltonian formulation). This equation provides a good approximation of the coorbital secular motion as long as the trajectories stay outside of the Hill sphere (here a circle of radius $r_0(\mu/3)^{1/3}$ whose center is the body of mass $m_1$). According to \cite{Kevorkian1970}, this condition imposes the parameter $\rho$ not to be very large with respect to unity. The system associated with the differential equation (\ref{eq:sol_orb_zeta}) possesses three fixed points corresponding to the two stable equilibria $L_4$ and $L_5$ for $(\zeta,\dot\zeta) = (\pm \pi/3,0)$, and to the unstable equilibrium $L_3$ when $(\zeta,\dot\zeta) = ( \pi,0)$. The collinear points $L_1$ and $L_2$ are not fixed points of the equation (\ref{eq:sol_orb_zeta}), but as the points lie on the boundary  of the Hill sphere, we do not consider this problem. In contrast to the real problem, in this one degree of freedom approximation given by the equation (\ref{eq:sol_orb_zeta}), the  stable and unstable manifolds associated with $L_3$ coincide (the system is integrable). The domain inside these manifolds is filled with tadpole orbits: periodic orbits surrounding $L_4$ or $L_5$  whose frequency varies from $\sqrt{27\mu}n/2$ in the neighborhood of the corresponding equilibrium to $0$ approaching the separatrix. Outside of the separatrices  lie the horseshoe orbits whose frequency ranges from zero on the separatrix to infinity when $\dot\zeta$ tends to infinity. However, the value of the frequency is bounded because when $\vert \dot\zeta\vert$ is large, the distance between the trajectory and the collision with the mass $m_1$ is small, that is contrary to our hypothesis that  the horseshoe trajectory must not enter the Hill sphere (see Eq.\ref{eq:sol_orb_zeta}). Therefore, it can be shown that  the frequency of the horseshoe trajectories  is bounded by a quantity of the order of $\smu n$, similar to the frequency in the tadpole regime.

The fast component of the motion is equal to $r_0\sqrt{\mu}\rho\cos(v_p + \zeta(t) + \psi) $ in the expression of $r$ and equal to $-2\sqrt{\mu} \rho\sin( v_p + \zeta(t) + \psi) $ for $v$. These two functions are quasiperiodic with frequencies $n$ and $\nu$ where $\nu = \cO(\smu)$ is the frequency of the periodic function $\zeta$ representing the secular motion. If we do not consider the function $\zeta$, the expressions of $r$ and $v$ are the same as those obtained by a first order expansion in eccentricity in the  Keplerian case, the eccentricity being equal to $\smu\rho$. This remark will be exploited below.

For the real trojan satellites of Saturn, formulas (\ref{eq:sol_orb_r}) to (\ref{eq:sol_orb_zeta}) provide a quite  rough approximation of their orbital motion. Indeed, even if we can assume that the motion of Dione (resp. Tethys) is not too much affected by their coorbital companions Helene  and Polydeuces (resp. Telesto and Calypso),  the motion of Dione and Tethys is not Keplerian. In particular, the $2:1$ mean motion resonances between Dione and Enceladus and  between Mimas and Tethys  generate long period variations of the orbital elements of these bodies (period of about $70.5$ years for Tethys and $11.1$ years for Dione \citep{ViDu1995}). Here, the effect of Saturn's oblateness is not directly taken into account. Its main impact on the orbital motion is to slightly modify the fundamental frequencies. 
As we derive these frequencies from the {\it Horizons} ephemerides that already include several spherical harmonics related to the shape of Saturn, the oblateness is implicitly considered. 
In addition, a second effect of the Saturn $J_2$ factor is to modify the triangular equilibrium configuration. Indeed, the equilateral triangle flattens slightly to give an isosceles triangle (see \citeauthor{SharmaRao1976}, \citeyear{SharmaRao1976}), but this modification is very small in regard to the other neglected perturbations.  Even if the orbital model that we consider here is not accurate enough to describe precisely the orbital motion of the studied satellites, its injection in the equations of the rotation is enough to describe the rotational libration with a better accuracy than the accuracy providing by the Cassini data \citep{TiThBu2009}.

The parameters used in this paper to model the orbits are displayed in the table \ref{tab:orb}.  $\mu$ is the mass of the main satellite associated with the coorbital in the three-body problem divided by the mass of Saturn.  The value of $\rho$ is fitted to the mean eccentricity of the satellite, derived from the ephemerides, because it controls the amplitude of the short time variations of $r$ and $v - nt$ therefore it acts like the mean eccentricity of the satellite. 
This crude approximation is sufficient to give a reasonable order of magnitude of the libration amplitude for the coorbital satellites (see Table \ref{tab:parameters}). The secular variations of $v$ and $r$ are represented by the function $\zeta$, solution of the differential equation (\ref{eq:sol_orb_zeta}). For the first four satellites of the table, this solution is chosen such a way that its period $2\pi/\nu$ ($\nu$ is given in the last column of the table)  is the same as the period of the libration around $L_4$ or $L_5$ (depending on the satellite) deduced from the \textit{Horizons} ephemerides \citep{Gioetal1996}.

For Janus and Epimetheus, the situation is more complicated because the long-term component of their true anomaly (the average of $v$ with respect to the orbital period, namely the $8$-years component) does not verify the equation (\ref{eq:sol_orb_zeta}). But, as it is shown in \cite{RoRaCa2011}, the average of the relative mean anomaly (difference between Janus' anomaly and the Epimetheus one) satisfies the differential equation (\ref{eq:sol_orb_zeta}). In order to describe the long term temporal evolution of $r$ and $v$ we use the expressions (\ref{eq:sol_orb_r}) and (\ref{eq:sol_orb_v})  again, but by replacing the function $\zeta$ by $m_E/(m_J+m_E)\zeta$ for Janus, and  $m_J/(m_J+m_E)\zeta$
for Epimetheus, where $\zeta$ represents the above mentioned long time variations of relative orbits of Janus and Epimetheus, is solution of the equation (\ref{eq:sol_orb_zeta}). More explanations concerning this point can be found in \cite{RoRaCa2011} and references therein. The minimum and maximum values of $\zeta$ are presented in  the fourth and fifth  columns of Tab. \ref{tab:orb}. For the orbital motion of Janus and Epimetheus, the variation in $\zeta$ is greater than 180 degrees resulting from the horseshoe shape orbits of these satellites. On the contrary, the variations in $\zeta$ are smaller for the satellites in tadpole orbits. 

\begin{table}[h]
\begin{center}
\begin{tabular}{ccccccc}
\hline
\text{satellite} &$\mu$ & $\rho$ & $\text{min} \zeta$ & $\text{max} \zeta$ & $n$ & $\nu$  \\ 
  & &  & (deg.) &  (deg.) & (rad/day)& ($10^{-3}$ rad/day)  \\ 
\hline
Polydeuces & $2.9\, 10^{-6}  $ &$11.5$ &$268$ & $321$ &$2.3$ &$7.93 $ \\
Helene         & $2.9\, 10^{-6}  $ &$4.6$   &$47$   &$77$     &$2.3$ &$8.18$ \\
Telesto         & $1.0\, 10^{-6}  $ &$0.6$   &$296$ &$304$  &$3.3$ &$9.03$ \\
Calypso       & $1.0\, 10^{-6}  $ &$0.8$   &$59$    & $61$   &$3.3$ &$9.02$\\
Janus           & $9.7\, 10^{-10}$&$218$  & $6$      &354      & $9.03$&$2.15$ \\
Epimetheus &$3.3\, 10^{-9}  $ &$171$  &$6$       &354      & $9.03$&$2.15$ \\
\hline
\end{tabular}
\end{center}
\caption{
Orbital parameters of the six coorbital satellites deduced from the \textit{Horizons} ephemerides:  $\mu = m_1/ (m_1+m_0)$, $\rho$ that appears in the expressions (\ref{eq:sol_orb_r}) and (\ref{eq:sol_orb_v}) is derived from the satellite's averaged eccentricity, $n$ is the mean motion and $\nu$ the coorbital frequency. The fourth and fifth columns give minimal and maximal values of the long-term component $\zeta$ satisfying the equation (\ref{eq:sol_orb_tau}).  }
\label{tab:orb}
\end{table}

\begin{figure}[htbp]
\begin{center}
\includegraphics[width=10cm]{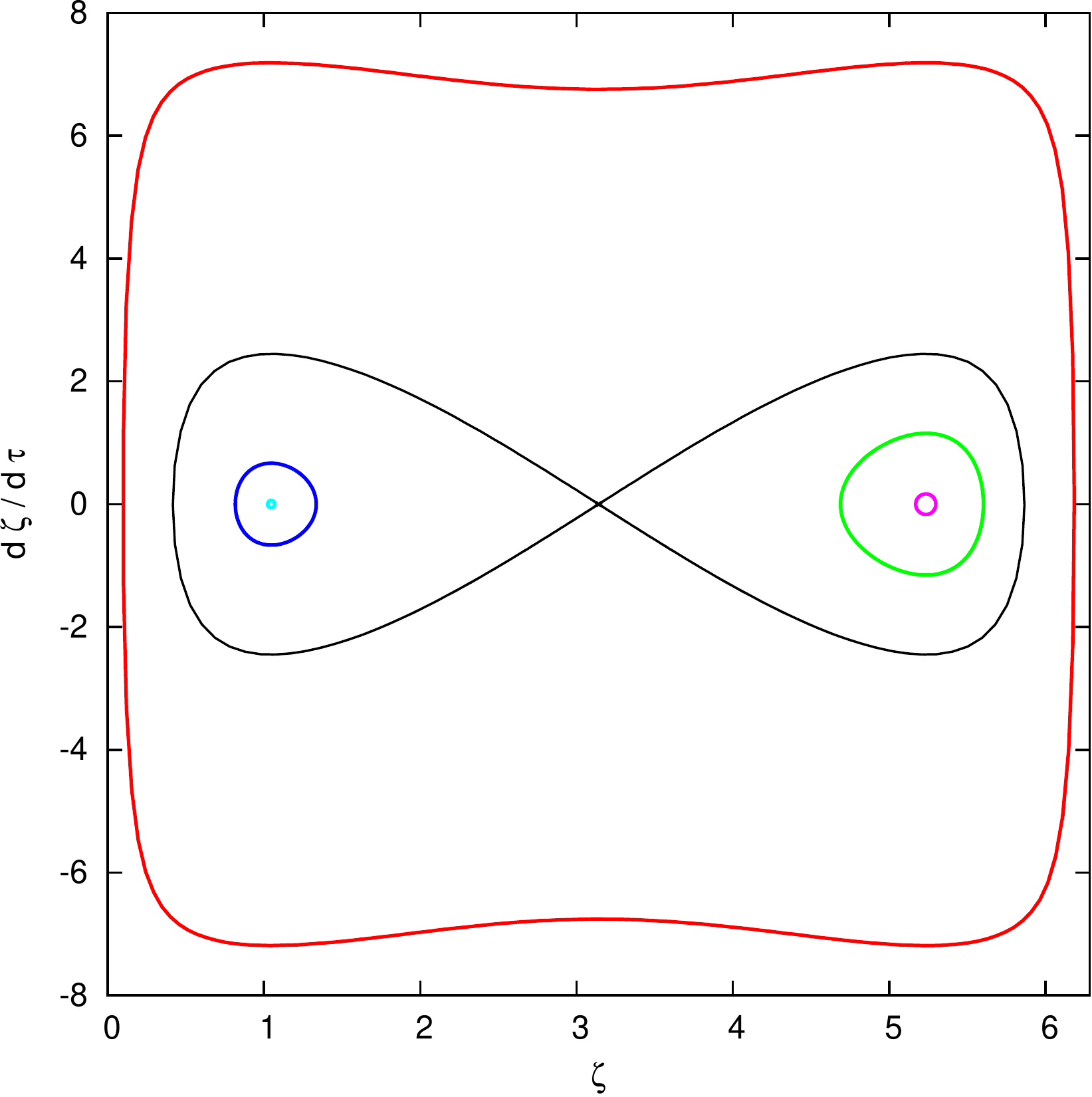}
\caption{Average orbit of the satellites modeled by the equation (\ref{eq:sol_orb_zeta}) plotted in the plan $(\zeta,d\zeta/d \tau)$. The coordinates of $L_3$, $L_4$ and $L_5$ are respectively $(\pi,0)$, $(\pi/3,0)$ and  $(5\pi/3,0)$. The two tadpole orbits surrounding $L_4$ are associated with Calypso (smallest one) and Helene, while the ones around $L_5$ corresponds to Telesto (smallest one) and Polydeuces.  The outermost orbit describes the average motion of Epimetheus with respect to Janus.}
\label{fig:orb}
\end{center}
\end{figure}

In order to display the average orbits of the 6 coorbital satellites on the same plot,  different time-scale is used for each body. Rather than $t$, we use the time $\tau = \smu n t$ in such a way that the differential equation (\ref{eq:sol_orb_zeta}) becomes free from any parameter, that is to say:
\be
 \der{^2\zeta}{\, \tau^2}   =  - 3\left(
               1 - (2-2\cos \zeta)^{-3/2}\right)\sin \zeta \label{eq:sol_orb_tau}.
\ee
This rescaling allows us to compare these orbits in the plan $(\zeta,d\zeta/d \tau)$ and to add the separatrix associated with $L_3$ (black curve) which discriminates the tadpole orbits surrounding $L_4$ and $L_5$ from the horseshoe orbits outside of the separatrix. For the four satellites in tadpole orbit (Polydeuces, Helene, Calypso, Telesto) the rescaling coefficient  $\smu n$ is derived from the table \ref{tab:orb}, while the initial conditions of (\ref{eq:sol_orb_tau}) at $\tau=0$ are $(\zeta_{max},0)$  where $\zeta_{max}$ is given in the fifth column of the table \ref{tab:orb}. As mentioned above, the individual orbit of Janus or Epimetheus is not directly related to the equation~(\ref{eq:sol_orb_zeta}), so we use their relative orbit (red in figure \ref{fig:orb}). This orbit is the integral curve of (\ref{eq:sol_orb_tau}) with $\tau = \sqrt{(m_J+m_E)/ (m_S + m_J+m_E)} n t$ which  includes the point of coordinates $(\zeta_{max}, 0)$.

\section{A Hamiltonian formulation of the rotation }
\label{sec:rotation}
\subsection{A simple quasiperiodic model}
The dynamical equation governing the physical libration is inferred from 
the angular momentum balance equation between the body's inertia and the gravitational torque exerted by Saturn projected onto the equatorial plane of the body. These equations have  the following form (see e.g. \cite{RoRaCa2011}):
\be
 \ddot \theta   +  \frac{\sigma^2}{2}\left( \frac{r_0}{r}\right)^3\sin2(\theta - v) = 0 , \text{ with } \sigma^2 = 3\frac{G m_0}{ r_0^3}\frac{(B-A)}{C},
 \label{eq:rot_theta}
\ee
where $\theta$ is the rotational angle, $G$ the gravitational constant, $m_0$ the mass of Saturn, $A,B,C$ are the principal moments of inertia such as $A \leq B \leq C$, and $(r,r_0,v)$ are the orbital parameters defined in the previous section. The frequency $\sigma$ is constant and corresponds to the frequency of libration at the equilibrium point. $\sigma$ depends on the difference $B-A$ and for a spherical body, or an axisymetric body along the north-south direction, it is equal to zero.
By introducing the angular distance  $x=\theta - v$ between the long axis of the satellite and the direction satellite-planet, we obtain the equation
\be
 \ddot x   + \frac{\sigma^2}{2}\left( \frac{r_0}{r}\right)^3\sin2x = -\ddot v ,
   \label{eq:rot_phi_exact}
\ee
whose solutions describe the motion of the planet in the satellite's sky. 

Using the approximation of $v$ given by the relation (\ref{eq:sol_orb_v}), we get 
\be
 \ddot x   +\frac{\sigma^2}{2}\left( \frac{r_0}{r}\right)^3\sin2x = - 2\smu \rho n^2\sin(v_p + \zeta(t) + \psi) ,
  \label{eq:rot_phi}
\ee
where the terms of order $\mu$ are neglected.  As mentioned in section \ref{sec:orbit}, $r$ and  the right hand side of the equation (\ref{eq:rot_phi}) are quasiperiodic functions of frequencies $n$ and $\nu$, thus the differential equation (\ref{eq:rot_phi}) depends quasiperiodically on the time.

The mass ratio $\mu$ being small, we adopt a perturbative method. We begin by assuming that $\mu = 0$. The equation (\ref{eq:rot_phi}),  which
is now equivalent to the one representing the motion of the pendulum, possesses a stable fixed point for  $\dot x = x = 0$ surrounded by periodic trajectories whose frequencies are close to $\sigma$ in a small neighborhood of the fixed point.
Now, increase the value of $\mu$ corresponds to perturb the pendulum equation in a quasiperiodic manner.
If the size of the perturbation (governed by $\mu$) is small enough, and under other  suitable conditions excluding some resonances between the frequencies $\sigma$, $n$ and $\nu$, the fixed point of the unperturbed equation is replaced by  a quasiperiodic solution of frequencies $(n,\nu)$ around which revolves quasiperiodic orbits possessing  $3$ frequencies (see \cite{JoSi1996}). The frequencies of the  central orbit are the two forcing frequencies $n$ and $\nu$, while the other orbits possess an additional "free" or "proper" frequency (close to $\sigma$) which depends on their initial conditions.  
The perturbed systems thus possesses  a one parameter family of 2-frequencies quasiperiodic orbits which end in  $x=\dot x=0$ when $\mu$ tends to  zero.

These forced solutions play a major role in the rotational dynamics of the close satellites. Indeed, in presence of dissipative phenomena, the forced solution will be a stable equilibrium (quasiperiodic attractor) towards which  most of the  trajectories will converge (see \cite{CeCh2008}).  
 It is worth mentioning that the quasiperiodic attractor of the dissipative case and the quasiperiodic forced solution of the conservative case are not the same trajectories (see \citeauthor{LandauLifchitz1960}, \citeyear{LandauLifchitz1960}, chap. 28). But their frequencies, which are the orbital frequencies, are independent from the dissipation and consequently are equal. Let us also mention that for small dissipations, the two trajectories are very close, and that the dissipative limit cycle tends to the conservative forced orbit when the dissipation vanishes. For these reasons, we will devote the rest of this article to the study of the dynamics in the neighborhood of the forced quasiperiodic orbit. 

\subsection{Hamiltonian formulation of the problem}
\label{sec:Ham}
\subsubsection{Autonomous Hamiltonian system}

In order to adopt an autonomous Hamiltonian formulation which takes into account the different time-scales, let us denote:  $\lam_1 = v_p + \psi$ and $\lam_2 = \smu n t$.  The solution $\zeta$ of the equation (\ref{eq:sol_orb_zeta}) can be rewritten as:
\be
\zeta(t)  =  \zeta(\lam_2{\smu}^{-1}n^{-1}) = \hzeta(\lam_2)
\ee
and its time derivative as:
\be
\der{\zeta}{t}(t) = \smu n  \hzeta'(\lam_2), \quad\text{with} \quad  \hzeta'(\lam_2) = \der{ \hzeta}{\lam_2}(\lam_2) .
\ee

It turns out that:
\be
\begin{split}
&\left(\frac{r_0}{r(t)}\right)^3 = 1 + \smu h(\lam_1,\lam_2) +\cO(\mu),\quad \text{where}\,\,  \\
&h(\lam_1,\lam_2) =  2\hzeta'(\lam_2) - 3\rho\cos u, \quad\text{with} \quad u = \lam_1+\hzeta(\lam_2)
\end{split}
\ee
and
\be
\begin{split}
&\dot v(t)  = n\left( 1 + \smu g(\lam_1,\lam_2)  \right) +\cO(\mu),\quad \text{where}\,\,  \\
&g(\lam_1,\lam_2) =  \hzeta'(\lam_2)  -  2\rho\cos u .
\end{split}
\ee
Finally, using the conjugated variables $(x,X) = (\theta - v(t), \dot\theta -n)$,  $(\lam_1,\Lam_1)$ and $(\lam_2,\Lam_2)$, the equation (\ref{eq:rot_phi}) is equivalent to the canonical equations associated with the Hamiltonian $H$ given by:
\be
H =  n\Lambda_1 + \smu n\Lambda_2 + \frac{X^2}{2} -\frac{\sigma^2}{4}\left(1 +  \smu h(\lam_1,\lam_2)\right)\cos 2x  -  n \smu g(\lam_1,\lam_2)X .
\ee
As we are only looking for the solutions in a neighborhood of zero, we expand the term $\cos 2x$ at order $4$. 
Using  the variables $(y,Y) = (\sigma^{1/2}x,\sigma^{-1/2}X)$, the expansion of the previous Hamiltonian (that we still denote by $H$) reads:
\be
\begin{split}
H =\,\,   & n\Lambda_1  + \sigma \frac{y^2 + Y^2}{2} -\frac{y^4}{6}   \\
&+ \smu \left(
              n\Lam_2 + \left( \sigma \frac{y^2}{2} - \frac{y^4}{6}\right) h(\lam_1,\lam_2)  - n\sqrt{\sigma}g(\lam_1,\lam_2)Y
             \right).
 \label{eq:H}
 \end{split}
\ee
In this expression the terms depending only on $\lam_1$ and $\lam_2$ are omitted. 

An usual way to  study  the dynamics for small values of $y$ and $Y$ is to reduce the Hamiltonian  (\ref{eq:H}) to a normal form (see e.g. \cite{ArKoNei2006}). This will be done in two consecutive steps. The first one, performed in Section \ref{sec:forceQP} leads to get rid of the linear terms in $Y$, that is to eliminate $g(\lam_1,\lam_2)Y$ using a new set of canonical coordinates. This is equivalent to bring back the force trajectory to the origin of the Cartesian coordinates. The second step, which is described in Section \ref {sec:nonres} consists in the replacement of the Cartesian coordinates by polar symplectic ones before averaging  the obtained expression.

\subsubsection{The quasiperiodic forced trajectory}
\label{sec:forceQP}

The suppression  of the term $g(\lam_1,\lam_2)Y$ in the expression  (\ref{eq:H})  is performed by introducing a new coordinate system  $(y',Y',\lam_1',\Lam_1',\lam_2',\Lam_2')$ linked to the coordinates $(y,Y,\lam_1,\Lam_1,\lam_2,\Lam_2)$, by the relations\footnote{Let us notice that the transformation (\ref{eq:transcan}) is canonical only when the terms of order two and more in $\smu$ are neglected}:
\be
y = y'- \dron{F}{Y'}, Y = Y' + \dron{F}{y'}, \lam_j = \lam_j '- \dron{F}{\Lam_j '}, \Lam_j  = \Lam_j ' + \dron{F}{\lam_j '},
\label{eq:transcan}
\ee
where the generating function $F$ is a solution of the partial differential equation:
\be
n\dron{F}{\lam_1'} + \sigma
   \left(
   Y'\dron{F}{y'} - y'\dron{F}{Y'}
   \right)
  = n\sqrt{\sigma}\smu g(\lam'_1,\lam'_2)Y' .
  \ee
We choose here the solution:
\be
F = n\sqrt{\sigma}\smu
\left(
\frac{\hzeta'(\lam'_2)}{\sigma} y'  -2\rho\frac{\sigma y' \cos(\lam'_1+\hzeta(\lam'_2)) - nY'\sin(\lam'_1+\hzeta(\lam'_2)) }{\sigma^2-n^2}
\right) .
\label{eq:generes}
\ee
This solution is valid only if $\sigma\neq n$, and more generally when $\vert\sigma - n\vert$ is not too small, in other words when the system is far enough from the 1:1 secondary resonance between the libration frequency $\sigma$ and the mean motion $n$.

In the coordinate system $(y',Y',\lam_1',\Lam_1',\lam_2',\Lam_2')$ the Hamiltonian (that we denote $H$ again)  becomes:
\be
H =  n\Lambda'_1  + \sigma \frac{y'^2 + Y'^2}{2} -\frac{y'^4}{ 6}      
+ \smu \left(
              n\Lam'_2 + \left( \sigma \frac{y'^2}{2} - \frac{y'^4}{6}\right) h(\lam'_1,\lam'_2)  
             \right).
 \label{eq:H'}
\ee

It is now clear that the trajectory 
\be
(y'(t), Y'(t), \lam'_1(t), \lam'_2(t)) = (0, 0, nt + {\lam'_1}^{(0)}, \smu nt +{\lam'}_2^{(0)} )
\ee
is a quasiperiodic solution of  (\ref{eq:H'}). This trajectory depending only on the orbital frequencies is the forced orbit which  reads in original coordinates: 
\be
x(t) = \theta(t) - v(t) = -2\rho\smu \frac{n^2}{\sigma^2-n^2}\sin(\lam'_1 + \hzeta(\lam'_2)).
\label{eq:sol_force}
\ee
The expression of the forced trajectory in a reference frame rotating with the angular velocity $n$ reads:
\be
\theta(t) -v_p(t) = \zeta(t)  -2\rho\smu \frac{\sigma^2}{\sigma^2-n^2}\sin(\lam'_1 + \hzeta(\lam'_2)).
\label{eq:sol_force_rot}
\ee
The solution is the sum of two terms of different nature: a long-period term, which amplitude is about $1^{\circ}$ for Calypso and can reach $135^{\circ}$ for Epimetheus \citep{RoRaCa2011}, and a term at the orbital period whose amplitude, given in Table \ref{tab:parameters}, does not exceed 
$9^{\circ}$. 
The amplitude of this libration is very similar to the one obtained for a satellite in Keplerian motion. If we equate the term $\rho\smu$ with the orbital eccentricity, we recover the "Keplerian" amplitude expression but the phase has a very different behavior. In contrast with the Keplerian case, the phase is a quasiperiodic function of the time and the solution is close to the solution obtained by adiabatic invariant method developed in \cite{RoRaCa2011}. 

\subsection{The non-resonant case}
\label{sec:nonres}

In order to proceed to the reduction of the Hamiltonian~(\ref{eq:H'}), let us introduce the symplectic polar coordinates 
\be
(y',Y') = (\sqrt{2W}\sin w,\sqrt{2W}\cos w).
\ee
In these variables, the Hamiltonian~(\ref{eq:H'}) can be split in three components: the unperturbed part $H_0$ given by  \be
H_0 = n\Lam'_1 + \sigma W, 
   \label{eq:H0}
\ee
the averaged part of the perturbation $\bH_1$ which reads
\be
\bH_1 = n\smu\Lam'_2 - \frac{W^2}{4} + \smu
    \left( 
     \sigma W - \frac{W^2}{2}
    \right) \hzeta'(\lam'_2), 
\ee
 and its remainder  $\tH_1$ given by 
\be
\begin{split}
\tH_1 =  &   \left( 
      -\smu\sigma W\hzeta' + (1 + 2 \smu\hzeta')\frac{W^2}{3}
      \right)C_{0,2}    - (1 + 2 \smu\hzeta')\frac{W^2}{12}C_{0,4} \\
  & + \smu\rho( \frac{3\sigma}{2} W - W^2) C_{1,2}  
   + \frac{\smu\rho W^2}{4}  C_{1,4} 
   - 3\smu\rho ( \frac{\sigma}{2} W - \frac{W^2}{4}) C_{1,0} ,
   \label{eq:Htilde}
\end{split}
\ee
with 
\be
2C_{p,q} = \cos(pu + qw) + \cos(pu - qw) = 2\cos pu\cos qw 
\label{eq:def_C}
\ee
where $u = \lam'_1 + \hzeta(\lam'_2)$.

These expressions show that the angles $w$ and $\lam'_1$ vary rapidly ($\dot w =\cO(\sigma)$ and $\dot \lam'_1 = \cO(n)$) while $\lam'_2$ is a slow angle ($\dot \lam'_2 = \cO(\smu n)$). Consequently, an usual way to find the approximate solutions of the previous Hamiltonian system is to  consider its average   with respect to the fast angles $\lam_1$ and $w$.

If, as it is the case in the previous section,  the terms of order $\mu$ and more are neglected, and if the term $\cos 2x $ is expanded in Taylor series up to degree $2p$ in $x$, the resonances that occur during the first order averaging process of the Hamiltonian (\ref{eq:Htilde}) are  $n + k\sigma =0$, $k$ being an integer satisfying $\vert k\vert \leq 2p$.

As mentioned above, the Hamiltonian $H$ can be reduced to its average with respect to the fast angles $\lam'_1$ and $w$  given by the expression:
\be
K = n\Lam'_1 + \sigma W
+  n\smu\Lam'_2 - \frac{W^2}{4} + \smu
    \left( 
     \sigma W - \frac{W^2}{2}
    \right) \hzeta'(\lam'_2),
\ee
where  the terms which are negligible with respect to $\smu$ and $W^2$ (when $\mu$ and $W$ tend to zero) are omitted.
The transformation linking $H$ to $K$ is approximated by the expression:
\be
z =  \bar z - \dron{F_a}{\bar  Z}, \quad  Z =  \bar Z + \dron{F_a}{\bar  z},
\ee
where  $(z,Z)$ is a couple of conjugated coordinates  belonging to  $\{(\lam'_1,\Lam'_1),(\lam'_2,\Lam'_2),(w,W)\}$,  $(\bar z,\bar Z)$ are the average variables,
and $F_a$ the generating function of the transformation. For simplicity, the "bar" symbols will be omitted in the rest of this section. 

The function $F_a$ being a solution of the equation 
\be
n\dron{F_a}{\lam'_1} +  \sigma\dron{F_a}{w} + \tH_1 = 0,
\ee
we adopt here:
\be
 F_a = -\sum_{p,q} H_{p,q}(W,\lam'_2)S_{p,q},
\ee
with 
\be
\begin{split}
S_{p,q} &= \frac12\left( 
                              \dfrac{\sin(pu + qw)}{pn + q\sigma} + \frac{\sin(pu - qw)}{pn - q\sigma}
                   \right)  \quad \text{if } \, H_{p,q} \neq 0, \\
S_{p,q} &= 0  \quad \text{if } \, H_{p,q} = 0.
\end{split}
\ee

Solving the Hamiltonian system associated with $K$ allows to give the solutions of the problem in the original variables $(x,X)$. In particular, we have:

\be
\begin{split}
x(t) &=   -2\smu\rho \frac{n^2}{\sigma^2-n^2}\sin \bu
+ \sqrt{2\bW\sigma^{-1} }\sin\bw \\
 & +\sqrt{\frac{\mu}{2\sigma} \bW}  \left[
  \hzeta'(\lam_2)\sin \bu + 
  \frac{3\sigma\rho}{4n}  \left(\frac{n+4\sigma}{n+2\sigma}\sin( \bu +\bw) +\frac{n-4\sigma}{n-2\sigma}\sin( \bu -\bw)\right)
  \right],
\label{eq:sol_x}
\end{split}
\ee
where $\bW$ is a constant and the temporal expressions of the angles $\bu$ and $\bw$ are:
\be
\begin{split}
\bu(t) =\, & nt + \zeta(t) +\psi , \\
\bw(t) =\, & \bw_0 + \left(\sigma - \frac{\bW}{2}\right)t + \frac{\sigma -\bW}{n} \zeta(t).
\end{split}
\ee
When $\bW =0$, we find the forced trajectory (\ref{eq:sol_force}) whose fundamental frequencies are only the orbital frequencies $(n,\nu)$. When $\bW>0$, the trajectories which oscillate around the forced trajectory have one additional frequency that depends on the distance to the forced orbit $\bW$: $\bar{\sigma}(\bW) =  \sigma - \bW/2 + \cO(\bW^2)$.  
We note that, unlike the naive representation which consists in separating the solution of the motion in a forced  and a free component, the part of the trajectory which depends on the parameter $\bW$ does not contain only $\bar{\sigma}(\bW)$  but mixed term containing the three main frequencies.

\subsection{Secondary resonances}
\label{sec:secondres}

We have seen in Section \ref{sec:forceQP} that due to the presence of the denominator $\sigma^2-n^2$ the transformation defined by the relations (\ref{eq:transcan}) to (\ref{eq:generes}), which maps a neighborhood of the forced orbit onto a neighborhood of $y'=Y'=0$, is not defined when the system approaches the 1:1 secondary resonance. 
Similarly, the averaging performed in Section \ref{sec:nonres} makes sense only when $n + k\sigma \neq 0$.
Consequently we will see on two particular cases (the 1:1 and 2:1 secondary resonances) how the topology of the phase space is modified when secondary resonances occur. Such a development of secondary spin-orbit resonance has been done by \cite{Wisdom2004} in the case of Keplerian orbit and more specifically for the 3:1 resonance.

\subsubsection{The 1:1 secondary resonance}
\label{sec:res_1:1}

In order to study the dynamics in the neighborhood of 1:1 resonance between the frequency of libration and the orbital frequency, we start from the initial Hamiltonian (\ref{eq:H}). Using the new coordinate system   $(w,W,\lam'_j,\Lam'_j)$  defined by the relation $(y,Y,\lam_j,\Lam_j) = (\sqrt{2W}\sin w, \sqrt{2W}\cos w,\lam'_j,\Lam'_j)$ the Hamiltonian reads 
\be
H = H_0 + \bH_1 + \tH_1 + n\smu\sqrt{2W\sigma}\left( 2\rho C_{1,1}  - \hzeta'(\lam'_2) C_{0,1}\right),
\label{eq:H_complet}
\ee
where the notations are defined by  the formulas  (\ref{eq:H0}) to (\ref{eq:def_C}). This Hamiltonian presents an additional term with respect to the previous case, the last term, that represents the linear term in $Y$ in Hamiltonian (\ref{eq:H}).
 Then, we define a new set of canonical variables as:
\begin{eqnarray}
z\phantom{'_1} &= w - u = w - (\lam'_1 + \hzeta(\lam'_2)) ,                                                     &\quad Z\phantom{_1} = W,  \nonumber\\
 \vartheta_1 &= \lam'_1, \phantom{u = y - (\lam'_1 + \hzeta(\lam'_2))\,\,\,,}           &\quad  \Theta_1 = \Lam'_1 + W, \\
 \vartheta_2 &= \lam'_2,  \phantom{u = y - (\lam'_1 + \hzeta(\lam'_2))\,\,\,,}          &\quad \Theta_2 = \Lam'_2 + \hzeta'( \lam'_2)W, \nonumber
\end{eqnarray}
where $z$ and $ \vartheta_2$ are slow angles. After having averaged the Hamiltonian (\ref{eq:H_complet}) with respect to  the only fast angle $ \vartheta_1$, the Hamiltonian reads:
\be
\begin{split}
K_{1:1} =  & n\smu\bar\Theta_2 + (\sigma-n)\left(1 + \smu\hzeta'(\bar\vartheta_2)\right)\bZ  \\
                  &-\left(1 + 2\smu\hzeta'( \bar\vartheta_2)\right)\frac{\bZ^2}{4}  + n\sqrt{2\bZ\sigma}\smu\rho \cos \bz,
\end{split}
\ee
where $(\bz, \bZ, \bar\vartheta_2, \bar\varTheta_2)$ represent the averaged variables as in section \ref{sec:nonres}.

As the term $\smu\hzeta'( \vartheta_2)$ is always small with respect to the unity, it can be neglected in first approximation\footnote{If the Trojan satellite is at $L_4$ or $L_5$, then $\hzeta'( \vartheta_2) =0 \quad \, \forall t$.}. Consequently, we are left with the one degree of freedom  Hamiltonian:
\be
K_{1:1}^{(0)} =    (\sigma-n)\bZ  
                  -\frac{\bZ^2}{4}  + n\sqrt{2\bZ\sigma}\smu\rho \cos \bz,
\label{eq:Hres1:1}
\ee
which corresponds to the  classical problem known as the second fundamental model of resonance, or also Andoyer's model \citep{HenLe1983}.

The fixed points of the Hamiltonian system correspond to the slow component of the forced orbits (average with respect to the angle fast angle $\lam_1$). 
If we set $Q = \sqrt{2\bZ}$, the coordinates ($\bz,\bZ)$ of these fixed points are given by the positive roots of the polynomial equation:
\begin{eqnarray}
&P^-(Q)  = Q^3 - 4(\sigma-n)Q  - 4n\rho\sqrt{\mu\sigma} = 0 \, \text{ if }\, z = 0 ,\label{eq:polym}\\
&P^+(Q)  = Q^3 - 4(\sigma-n)Q  + 4n\rho\sqrt{\mu\sigma} = 0 \, \text{ if }\, z = \pi . \label{eq:polyp}
\end{eqnarray}
As $P^-(-Q) = -P^+(Q)$ , the roots of one of these polynomials are the opposite of the roots of the other one.

By fixing the parameters $\mu$, $\rho$, $n$ and vary the frequency $\sigma$ (in the neighborhood of $n$), there exists a critical value $\sigma_0$ such that for $\sigma < \sigma_0$ the average resonant system has only one fixed point, which is stable. Then, when $\sigma$ reaches the critical value $\sigma_0$ a pitchfork bifurcation with symmetry breaking (see e.g. \cite{LiLi1992}) occurs and gives rise to two new branches of fixed points. One of these new families contains stable fixed points while the other one possesses unstable equilibrium points. The bifurcation arises when the discriminant $\Delta$ of the polynomial involved in (\ref{eq:polym}) or (\ref{eq:polyp}) is equal to zero, that is 
\be
\Delta = 16\left( 
                    n^2\sigma\mu\rho^2 - \frac{16}{27}(\sigma  - n)^3  
                   \right) =0 ,
\ee
or 
\be
16\delta^3 - 27\mu\rho^2n^2\delta -  27\mu\rho^2n^3 =0\quad \text{with}\quad  \delta =\sigma  - n.
\ee
Since $\mu\rho^2 < 4$, this equation has only one real root $\delta_0$ given by:
\be
\begin{split}
\delta_0 &= \frac{3n}{4} (2\mu\rho^2)^{1/3}
    \left[
       \left( 1 + \sqrt{1-\frac{\mu\rho^2}{4} }\right)^{1/3} +
       \left( 1 -  \sqrt{1-\frac{\mu\rho^2}{4} }\right)^{1/3}
    \right] \\
    &= \frac{3n}{4} (4\mu\rho^2)^{1/3} +o(\mu^{1/3}) .
\end{split}
\label{eq:dl_delta}
\ee
In other words, the critical value of $\sigma$ is:
\be
\sigma_0 = n + \delta_0 = \left(1 + \frac{3}{4} (4\mu\rho^2)^{1/3}\right) n +o(\mu^{1/3}).
\label{eq:dl_deltabis}
\ee

Finally, the family of fixed points that exists for all values of $\sigma$ is given by $(\bz_1,\bZ_1) = (0,Q_1^2/2)$ where $Q_1$ is the positive root of the polynomial equation (\ref{eq:polym}) given by the expression
\be
Q_1 =  \left( 2n\rho\sqrt{\mu\sigma}  +\frac{\eta}{2}\sqrt{\vert\Delta\vert}\right)^{1/3}  +  \left( 2n\rho\sqrt{\mu\sigma}  -\frac{\eta}{2}\sqrt{\vert\Delta\vert}\right)^{1/3},
\ee
where $\eta =1$ if $\sigma <\sigma_0$ and $\eta =\sqrt{-1}$ if $\sigma \geq \sigma_0$.

The two others fixed points which coincide when $\sigma = \sigma_0$ and split for larger values are of the form $(\bz_p,\bZ_p) = (\pi,Q_p^2/2)$, where the $Q_p$'s ($p = 2,3$), which are the  two positive roots of $P^+$, can be calculated in the following way.
Let us define the complex number $u$ by the expression 
\be
 u =   -2n\rho\sqrt{\mu\sigma}  + \frac{\sqrt{-1}}{2}\sqrt{\vert\Delta\vert}.
\ee
As the real part of $u$ is negative and its imaginary part positive, $\phi$, the argument of $u$, is between $\pi/2$ and $\pi$. It turns out that the two positive roots of $P^+$ read:
\be
Q_2 = 4\sqrt{\frac{\sigma - n}{3}} \cos\frac{\phi}{3}, \quad Q_3 = 4\sqrt{\frac{\sigma - n}{3}} \cos\frac{\phi + 4\pi}{3} .
\ee

It is worth mentioning that, as $P^-(-Q) = -P^+(Q)$, the positive root of $P^-$ can also be read:
\be
 Q_1 = -4\sqrt{\frac{\sigma - n}{3}} \cos\frac{\phi + 2\pi}{3}\quad \text{for} \quad \sigma \geq \sigma_0.
\ee
The expression of the $Q_j$ are particularly simple when $\sigma = \sigma_0$. Indeed, as for this value of  $\sigma$ the argument $\phi$ is equal to $\pi$, we have
\be
Q_1 = 4\sqrt{\frac{\delta_0}{3}}, \quad Q_2 = Q_3 = 2\sqrt{\frac{\delta_0}{3}} .
\label{eq:Q_delta0}
\ee

Regarding the stability of these three equilibrium points, a straightforward  calculation shows that the characteristic polynomial of the linearized differential system in the neighborhood of $(\bz_j,\bZ_j)$ is given by:
\be
\cP_j(h) = h^2 +  \frac14 n\sqrt\sigma \smu\rho\frac{P'(Q_j)}{Q_j} ,
\label{eq:racines}
\ee
$P'$ being the first derivative of $P^-$ if $j=1$ and of $-P^+$ if $j=2,3$.

The polynomial function $P^-(Q)$ being increasing for all values of $\sigma$, its derivative is positive and the roots of $\cP_1$ are purely imaginary numbers. It turns out that the fixed point of coordinates $(0,\bZ_1)$ is linearly stable. 
The other family is more interesting. Indeed, when $\sigma=\sigma_0$, $Q_2= Q_3$ is a double root of $P^+$ and consequently its first derivative vanishes. This "double" equilibrium point is then degenerated (its eigenvalues are both equal to zero). As soon as $\sigma$ becomes larger that $\sigma_0$, the point associated with $Q_2$ becomes hyperbolic, while the one corresponding to $Q_3$ becomes elliptic. 
Indeed, the polynomial $P^+$ having three distinct roots, one negative and the two others $Q_2$ and $Q_3$ positive, the produce $\der{P^{+}}{Q}(Q_2) \der{P^{+}}{Q}(Q_3)$ is negative. As $Q_3\leq Q_2$,  $\der{P^{+}}{Q}(Q_2)>0$ and $\der{P^{+}}{Q}(Q_3)<0$, which proves the assertion stated above.

A straightforward computation, based on the determination of the generating function used to averaging  the Hamiltonian (\ref{eq:H_complet}) with respect to $\vartheta_1$, allows us to express the fixed points of (\ref{eq:Hres1:1}) in the original variable $(x,X)$. In particular, we have:
\be
x_j = \frac{Q_j}{\sqrt{\sigma}}\sin(\bz_j + \lam_1)  +   o\left(\smu^{\frac13 }\right), \quad X_j = \dot x_j +\cO(\smu) .
\label{eq:pf_x}
\ee
In the previous expression, the index $j$ is equal to $1$ if $\sigma \leq \sigma_0$ (only one fixed point) and to $1, 2$ or $3$ otherwise (three fixed points). 
When $\sigma$ is close to the bifurcation value $\sigma_0$, according to the expressions (\ref{eq:dl_deltabis}) and (\ref{eq:Q_delta0}), we have:
\be
x_j =  c_j\left(  2\smu\rho\right)^{\frac13} \sin(\bz_j + \lam_1)  + o(\smu^{\frac13}),
\ee
with $c_1 = 2$ and $c_2=c_3 = 1$. 
Consequently, the amplitude of the resonant case is much larger than in the non-resonant case, where the forced trajectory is of amplitude $\cO(\smu)$ (see formula (\ref{eq:sol_force}) and (\ref{eq:sol_x})). 

\begin{figure}[!h]
\begin{center}
\includegraphics[width=10cm]{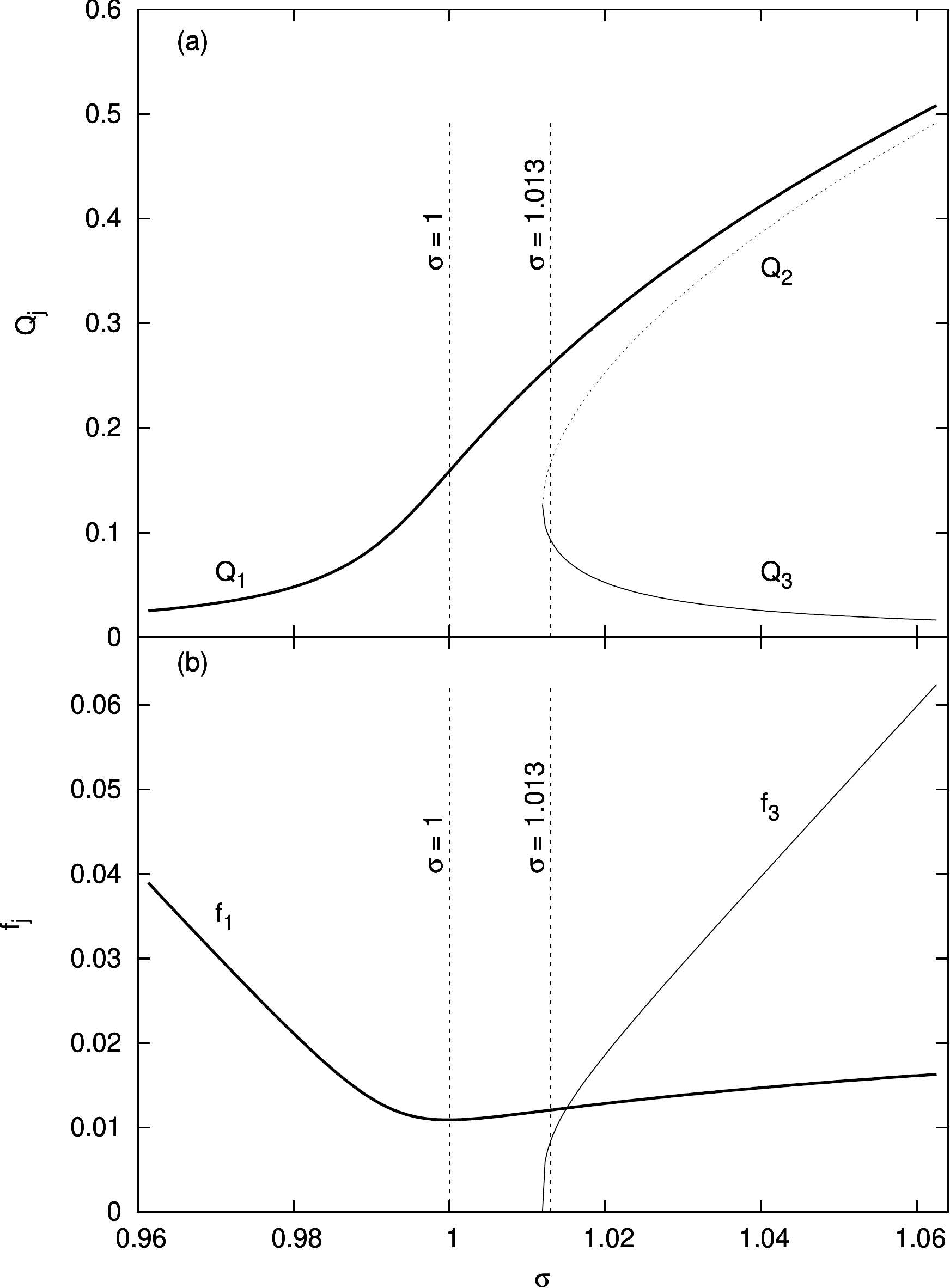}
\caption{Bifurcation inside the 1:1 secondary resonance. The upper graph displays the values of the $Q_j$ (Y-axis)  versus $\sigma$ (X-axis).  The quantities associated with the stable equilibria are represented with solid lines while dotted lines are associated with the unstable equilibrium. 
The frequencies of the libration around the two stable fixed points are plotted in the bottom graph. 
The two vertical dotted lines indicate the values of the parameter $\sigma$ used in the numerical simulations presented in Fig.~\ref{fig:res1_1}.}
\label{fig:res1_1_pf}
\end{center}
\end{figure}

Now we illustrate the analytical results by some numerical simulations. 
Until the end of this section, we fixed the free parameters of the model such that : $\mu = 1\times 10^{-6}$,  $n = 1$, $\rho = 1$,  $\psi= -\pi/3$ and $v_{p,0} = 0$. According to the expression (\ref{eq:dl_deltabis}) the critical value $\sigma$ is then equal to $\sigma_0 = 1.011953$.  

 Let us first illustrate the evolution of the fixed points of the Hamiltonian (\ref{eq:Hres1:1}) close to this bifurcation value. The quantities $Q_j$, which determine the location of the equilibrium points  through the formula (\ref{eq:pf_x}) are plotted in Fig \ref{fig:res1_1_pf}.a for $\sigma$ varying in the interval $(0.96,1.064)$ including $\sigma_0$. For $\sigma$ smaller than the critical value the phase space has only one equilibrium point defined by $Q_1$ (bold curve). In this domain, $Q_1$ increases since the distance of the fixed points from the origin is of order $\mu^{1/2}$  and of order $\mu^{1/3}$  when $\sigma$ is close to $\sigma_0$. After the bifurcation, this equilibrium point moves away from the origin.

 For $\sigma > \sigma_0$, the stable equilibrium associated with $Q_1$ persists while a new pair of fixed points appears at a distance of order $\mu^{1/3}$ from the origin: an unstable point related to $Q_2$ (dotted line) and a stable one associated with $Q_3$ (thin solid line). While the unstable point moves away from the origin as $\sigma$ increases, $Q_3$ decreases, reaching values of order  $\smu$ outside the resonance. In other words, the stable fixed point associated  with $Q_3$ plays the same role after  the 1:1 resonance than the one connected to $Q_1$ before this secondary resonance. 
The figure \ref{fig:res1_1_pf}.b shows the evolution of the frequencies $f_1$ and $f_3$ of the two elliptical (stable) fixed points  deduced from the equation ${\cal P}_j(\sqrt{-1}f_j) =0$, where the ${\cal P}_j$ are defined by the relation (\ref{eq:racines}). When $\sigma$ increases, the value of the frequency $f_1$ begins to decrease, reaches a minimum for $\sigma$ slightly greater than $1$, and then increases slowly.
The fixed point that is created at the bifurcation being degenerated the frequency $f_3$ (dotted curve) starts from zero and then increases rapidly along the stable branch.

\begin{figure}[!htb]
\begin{center}
\includegraphics[width=12cm,height=10cm]{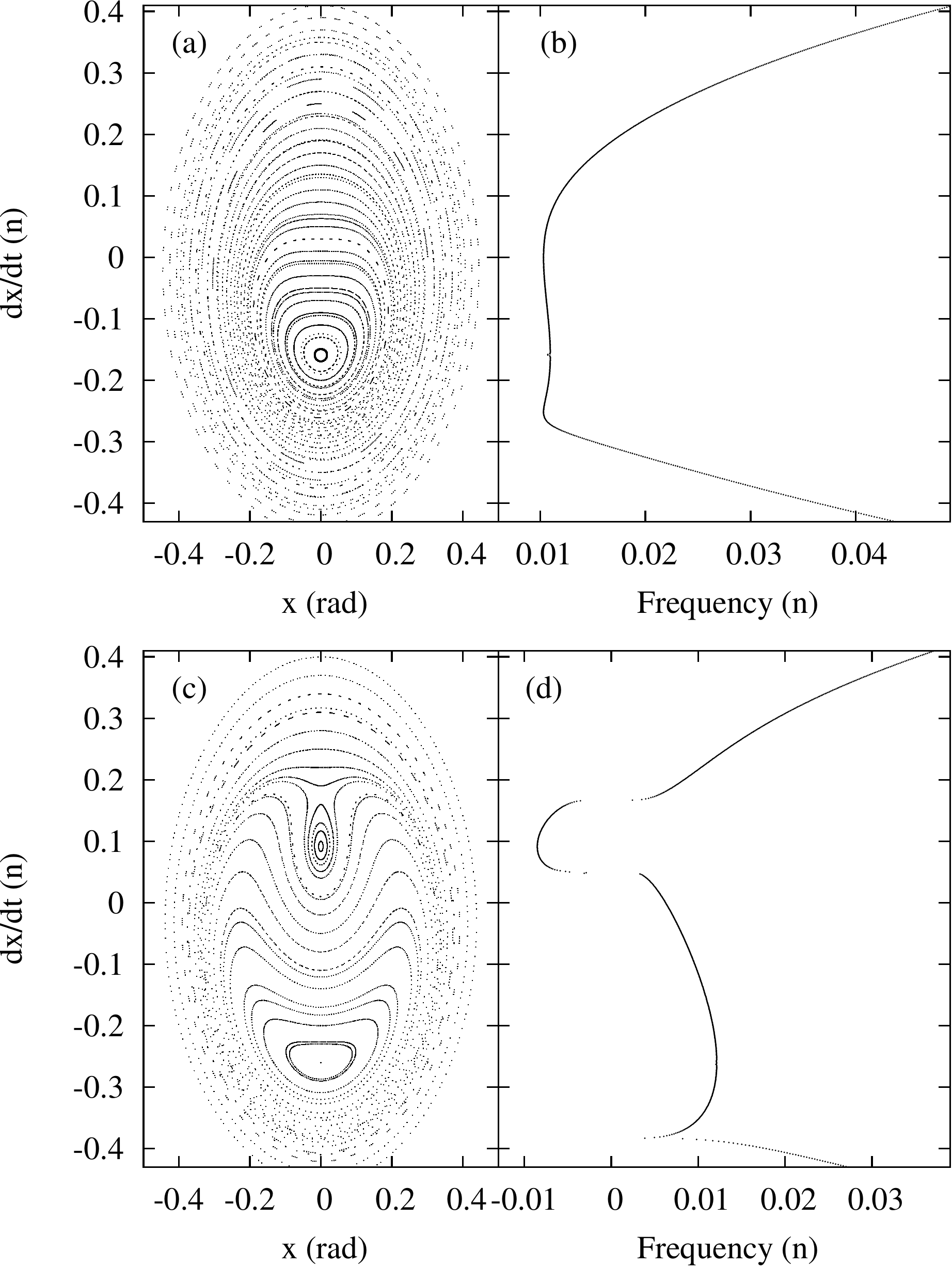}
\caption{
Phase portraits of the 1:1 secondary resonance in the coordinate system $(x,\dot x)$.  The figures (a) and (c) show the iterates of the first return map of the  unperturbed problem for two different values of $\sigma$ chosen in both side of the critical values $\sigma_0$: $\sigma = 1$ in figure (a) and $\sigma= 1.013$ in (c). The figures (b) and (d) represent the libration frequency $f(\dot x)$ for the corresponding values of $\sigma$.
The two gaps visible in figure (d) correspond to the crossing of the separatrices where a logarithmic singularity is encountered. The tangent to the frequency curve in this point is consequently horizontal.}
\label{fig:res1_1}
\end{center}
\end{figure}

In order to get more details regarding  the structure of the phase space in the neighborhood of the forced trajectory, we use Poincar\'e sections. To this end, we will not consider averaged equations, or expansions in a neighborhood of zero as we have done to carry out the analytical computations, but the complete initial differential equation (\ref{eq:rot_phi_exact}) where the temporal evolution of $v$, $r$ and $\zeta$ is deduced from  (\ref{eq:sol_orb_r}) to (\ref{eq:sol_orb_zeta}). 
As in the above analytical development, we have neglected the terms containing the function $\hzeta'$, the function $\zeta$ has to be constant along the integration. This can be done easily assuming that the satellite is at the equilibrium points $L_4$ or $L_5$, that is $\zeta = \hat\zeta = \pm \pi/3$ and $\dot\zeta = \hzeta' = 0$. 
Consequently, we are left with a one degree of freedom system depending periodically on the time (just like in the Keplerian case) and the phase portrait can be studied using the first return map defined by the flow of the system at time $2\pi/n$.  The global phase portrait of this problem is well known, it resembles to the one of the $\pi$-periodic pendulum weakly perturbed by a periodic time dependent function (see e.g. \cite{Wi1987A, Wisdom2004}).  Here we focus on a neighborhood of $x = \dot x = 0$ because we are interested in a local phenomenon that disturbs the forced orbit when $\sigma$ is close to the critical value $\sigma_0$.

The Figs. \ref{fig:res1_1}a and \ref{fig:res1_1}c show two different  phase portraits of the system in coordinates $(x, \dot x)$, restricted to a small area around the origin. 
The Fig.\ref{fig:res1_1}a is computed with $\sigma=1$  chosen below the critical value $\sigma_0 \approx 1.011953$ (Eq. \ref{eq:dl_deltabis} ), whereas on the bottom plot (c) $\sigma=1.013 $ is chosen slightly above the critical value $\sigma_0$. The phase space (a) shows only one elliptical fixed point (a periodic orbit) which is not centered on zero due to the non-zero value of $Q_1$. For the phase space (c), a second elliptical point appears at the position $Q_3$ because the system has crossed the bifurcation located at $\sigma_0$. Around these points, the trajectories rotate in two different directions and these two dynamical states are separated by the separatrix emerging from an hyperbolic point located just upper the $Q_3$ fixed point.

The Figs. \ref{fig:res1_1}b and \ref{fig:res1_1}d represent the graphs of the frequency map of the system for the two selected values of $\sigma$.
More precisely, assuming that the system is regular, the orbit of the first return map passing through the point of coordinates $(0, \dot x)$ is quasiperiodic and has two fundamental frequencies: $n$ and an independent frequency denoted  $f(\dot x)$. The frequency map is the map which associated with $\dot x$ the frequency $f(\dot x)$.
In practice, this frequency is evaluated numerically applying the frequency analysis developed by \cite{La90,La99}. If the dynamics of the studied system is regular (integrable system), the graph of the frequency map is almost everywhere perfectly smooth. On the contrary, its singularities or  lack of smoothness indicate the location of chaotic zones, or of hyperbolic domains associated with resonances. In addition, when a stable island associated with a resonance is crossed, the frequency $f(\dot x)$ remains constant. For further explanations and developments about the frequency map analysis, we refer the reader to \cite{La99} and references therein. 

The frequency $f(\dot x)$ versus $\dot x$ is plotted in the right column of the figure \ref{fig:res1_1}. 
In Fig. \ref{fig:res1_1}.b, where $\sigma = 1$, the curve seems  perfectly smooth, which indicates the global stability of the considered problem (if unstable regions exist, their size is microscopic at the scale of the figure). The frequency reaches a local maximum corresponding to the fixed point. Its value fits very well to the value predicted by the relation (\ref{eq:racines}). The two local minima are not directly related to critical points or singular structures of the phase space, but by increasing the values of $\sigma$, these minima will tend to zero. When $\sigma = \sigma_0$, one of these minima indicates the location of the fixed point $Q_2= Q_3$ and the other one, the location of its separatrix. 
Notice that, for $\sigma = 1$, the frequency $f(\dot x)$ is always greater that $0.01$, that will be of interest when the coorbital perturbation will be added to the model.  

For $\sigma = 1.013$ (Fig. \ref{fig:res1_1}.d), the frequency curve is globally smooth, but possesses three singularities associated with two gaps (close to $\dot x = 0.15$ and $0.18$) and a cusp (at $\dot x = -0.4$). These structures correspond to the crossing of the separatrices emerging from the unstable equilibrium. Between these singularities, the frequency reaches two extrema, the positive one, which coincides with the equilibrium located inside the big clockwise libration island, and the negative minimum associated with the fixed point lying inside the small counterclockwise libration island.  Here again, the agreement between the values of the extrema and the values predicted by $f_1$ and  $f_3$ is better than  $1\%$.

\begin{figure}[!htb]
\begin{center}
\includegraphics[width=12cm]{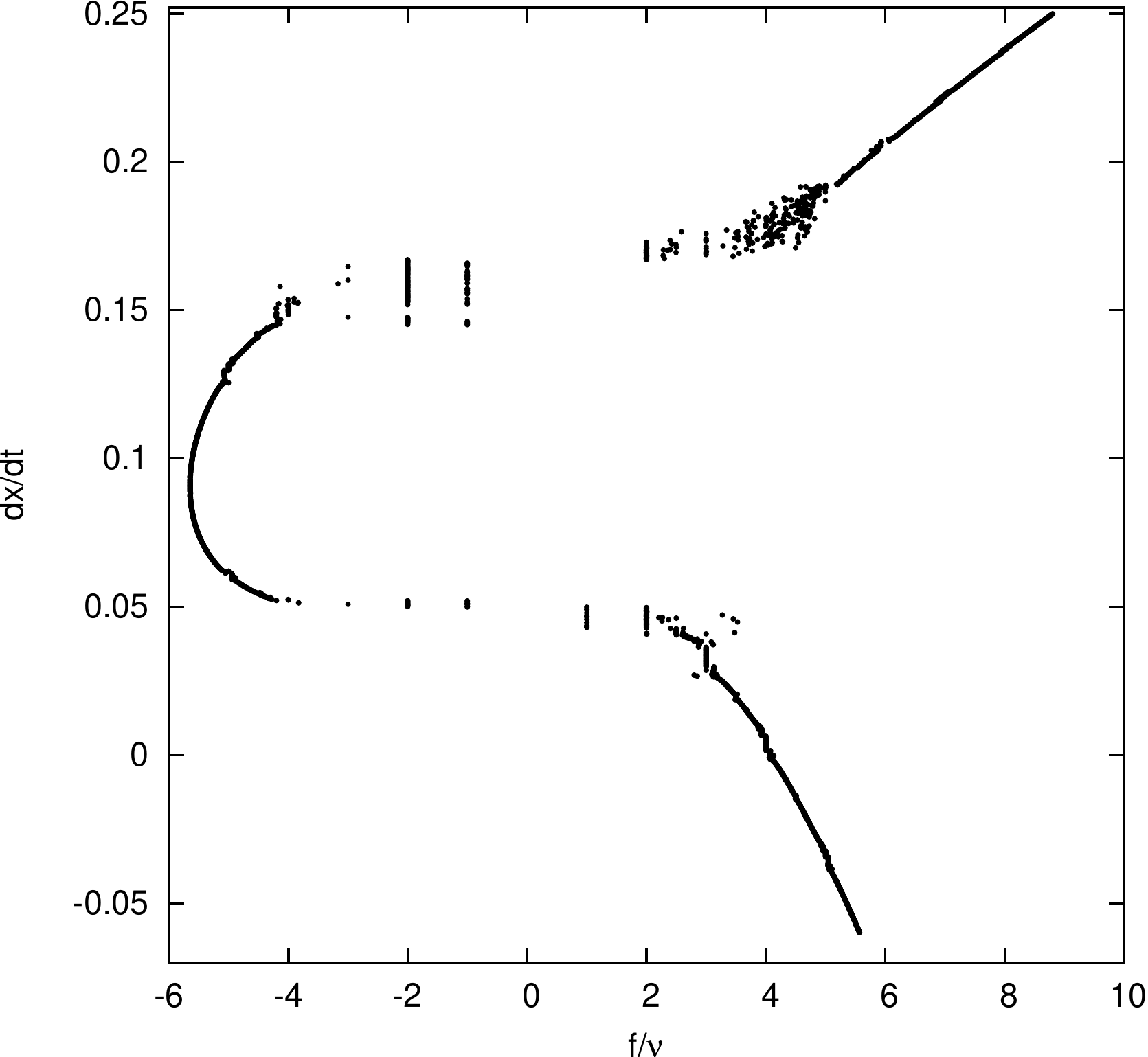}
\caption{
Chaotic behavior of the libration frequency of the perturbed problem:  enlargement of the counterclockwise libration island. The X-axis represents the ratio of the frequencies $f(\dot x)/\nu$, thus the abscissa of the main resonance  integer (or half-integer).  The values of $\dot x$ are on the Y-axis. This figure highlights the overlap of the lower order resonances and its associated chaotic area. The regions where $f(\dot x)$ is constant (vertical segments) are associated with the resonances $f/\nu = p \in \NN$. }
\label{fig:res1_1zoom}
\end{center}
\end{figure}

Let us now study the effect of the coorbital perturbation on the rotation inside the 1:1 secondary resonance. As the perturbed problem has three degrees of freedom, we cannot visualize its dynamics using Poincar\'e sections. But we can still study the frequency map of the perturbed problem. Here the situation is very different from the non-resonant case studies in section \ref{sec:nonres}. Indeed, in the non-resonant case, the libration frequency is close to $\sigma$. It turns out that the ratio of this frequency by the coorbital frequency $\nu$ is of order $1/\smu$. Then, the possible resonances between these frequencies are of high order and their dynamical influence is negligible.  But, as in the resonant case, the libration frequency can reach null values, so we can expect to observe interesting phenomena that we detail below.

As the size of the perturbation increases with the amplitude of the variations of $\zeta$, a large amplitude tadpole orbit has been chosen to introduce the perturbation in our simulation. Its  initial conditions are  $\zeta(0) = \pi/3, \dot\zeta(0) = 0.0024$. Along this orbit, $\zeta$ varies between  $24.3^{\circ}$ and $162.5^{\circ}$ and the coorbital frequency is equal to  $\nu = 1.5085 \, 10^{-3}$. 

In the  case of $\sigma = 1$, the frequency curve of the perturbed problem (not presented in the paper) is extremely similar to the one of the unperturbed problem drawn in Fig. \ref{fig:res1_1}.b. This shows that the addition of the perturbation does not generate any significant instability. This is probably due to the fact that the minimum of the libration frequency being of about $0.0103$, the lower order reachable resonance is given by $f = 7\nu$, and the system is in some sense protected from the destabilizing influence of the perturbation. Consequently, the chaotic zones resulting from these resonances are extremely small.   

On the contrary, the situation is much more interesting when $\sigma = 1.013$ for at least two reasons. First, due to the presence of a hyperbolic fixed point, very thin chaotic regions exist near the separatrices of the unperturbed problem. Then we expect that, adding one degree of freedom, the size of these chaotic region increases.  The second reason lies on the fact that, as shown in Fig \ref{fig:res1_1}.d, the frequency $f$ approaches zero, making possible resonances of low order between $f$  and $\nu$.
The expected phenomena are observed in Fig \ref{fig:res1_1zoom} which presents an enlargement of the graph of the perturbed  frequency map. 
The global shape of the frequency curve is similar to the case of the unperturbed problem (Fig. \ref{fig:res1_1}.d), but new structures are clearly visible.  This curve presents two very different regions. The first one is composed of three pieces of the curve: the part lying above the line  $\dot x =0.196$, the part  below $\dot x = 0.025$ and the one located in the left side of the vertical line $f/\nu = -4$. In this first domain, the curve is very smooth except for some regions marked by discontinuities generated by resonances between $f(\dot x)$ and $\nu$. But they are isolated and their associated chaotic areas are not observable, even at this scale.  

The second region includes what remains of the two (quasi) horizontal branches of Fig. \ref{fig:res1_1}.d marking the crossing of the separatrices of the unperturbed problem. The regular branches are replaced by vertical segments, whose abscissa is integer,  and by regions where the points are not well arranged.  These structures are mainly generated by the presence of low order resonances   $f = k\nu$, the integer $k$ satisfying  $-4\leq h \leq 5$,  and of their overlapping. Although these phenomena are clearly visible, they are confined to the vicinity of the separatrices of the unperturbed problem. In the parameter space that we explored, this should not significantly affect the rotation of the satellite. By extension, in the case of the coorbitals, with the parameters listed in Tables \ref{tab:orb} and \ref{tab:parameters}, the resonances should stay confined in very limited phase space.

\subsubsection{The 2:1 secondary resonance}

According to Section \ref{sec:nonres}, the invariant set $(y'(t), Y'(t), \lam'_1(t), \lam'_2(t)) = (0, 0, nt + {\lam'_1}^{(0)}, \\ \smu nt +{\lam'}_2^{(0)} )$ corresponds to the forced trajectory which is elliptic (stable) in absence of secondary resonances. As its normal frequency is close to $\sigma$, a doubling period like bifurcation arises when $\sigma$ is close to $n/2$. In order to study this phenomenon, we introduce the canonical coordinates:  
\begin{eqnarray}
z\phantom{'_1} &= 2w - u = 2w - (\lam'_1 + \hzeta(\lam'_2)) ,                                                     &\quad Z\phantom{'_1} = W/2 , \nonumber\\
 \vartheta_1 &= \lam'_1, \phantom{u = 2w - (\lam'_1 + \hzeta(\lam'_2))\,\,\,\,,}           &\quad  \Theta_1 = \Lam'_1 + W/2 ,\\
\vartheta_2 &= \lam'_2,  \phantom{u = 2w - (\lam'_1 + \hzeta(\lam'_2))\,\,\,\, ,}          &\quad \Theta_2 = \Lam'_2 + \hzeta'(\lam'_2)W/2 , \nonumber
\end{eqnarray}
such that $z$ and $\vartheta_2$ are two slow angles.
After the substitution of  $(w, W, \lam'_j, \Lam'_j)$ by the new coordinates $(z, Z, \vartheta_j, \Theta_j)$ in the Hamiltonian  $H$ defined by the relations (\ref{eq:H0}) to (\ref{eq:Htilde}), followed by its averaging with respect to $\vartheta_1$, the resonant Hamiltonian reads\footnote{In order to simplify the notations, the bars symbols over the variables are omitted in the expression of the averaged Hamiltonian.}:
\be
\begin{split}
K_{2:1} = (2\sigma -n) Z - Z^2 +\frac{\smu\rho}{2}\left( 3\sigma Z - 4Z^2 \right)\cos z \\
+\smu n\Theta_2 +
\smu\left(
 2\sigma -n -2Z 
 \right) Z \hzeta'(\vartheta_2) .
\label{eq:Hres21}
\end{split}
\ee
As in Section \ref{sec:res_1:1}, in a first step, we can focus on the first line of the previous expression, neglecting the terms depending on $(\vartheta_2,\Theta_2)$. 
The topology of the phase space of this Hamiltonian system whose Hamiltonian is denoted by $K_{2:1}^{(0)}$ bifurcates for two different values of the parameter $\sigma$: 
\be
\sigma_0 = \frac{n}{2}\left( 1 - \frac32\smu\rho \right)^{-1},\quad\text{and }\quad \sigma_{\pi} = \frac{n}{2}\left( 1 + \frac32\smu\rho \right)^{-1} .
\ee
When $\sigma  < \sigma_0$ the system associated with $K_{2:1}^{(0)}$ does not have any fixed point, except the point $Z=0$, which corresponds to the singular half-line of the polar coordinates.  In other words, in cartesian coordinates $(\sqrt{2Z}\sin z ,\sqrt{2Z}\cos z)$, the origin is a stable fixed point, and the topology of the phase space is similar to the non resonant case. 
When the first critical value $\sigma_0$ is reached, the origin (or the half-line $Z=0$) loses its stability given rise to a new elliptical fixed points of coordinates $(z_0,Z_0)$ with
\be
z_0 = 0,\quad  Z_0 = \frac{2(2\sigma -n) + 3\smu\rho\sigma}{4(1 +  2\smu\rho)} .
\ee
From the second critical value $\sigma_{\pi}$, the central equilibrium point recovers its stability while a new unstable fixed point of coordinates $(z_{\pi},Z_{\pi})$,  where
\be
z_{\pi} = \pi,\quad  Z_{\pi} = \frac{2(2\sigma -n) - 3\smu\rho\sigma}{4(1 -  2\smu\rho)},
\ee
 emerges from $Z=0$. Its separatrix surrounds the  point $(z_0,Z_0)$  which becomes stable again.
This kind of bifurcation is described by Murray and Dermott (1999, Chap. 8, Fig. 8.12)  for a different situation corresponding to an internal second-order  mean motion resonance in the restricted three body problem.  One gets the same phase portrait as in the book using a  coordinate system like $(Z\cos(z/2),Z\sin(z/2))$.

As it is the case for the 1:1 secondary resonance studied in Section \ref{sec:res_1:1}, if the value of the parameter $\sigma$ exceeds  $\sigma_0$,  the frequency can approach  zero.  If we take into account the orbital perturbation, this will give rise to resonances with the  frequency $\nu$, generating wide chaotic regions in the neighborhood of the separatrices of the unperturbed problem. But the goal of this paper is not to describe these local phenomena.

\section{Discussion and Conclusion}  
\label{sec:discu}

In previous sections, we have described the dynamical behavior of the rotational motion of the coorbital satellites. Notably, we show that three parameters control the dynamics (i) the frequency $\sigma$ related to the dynamical figure of the satellite, (ii) the departure from the Lagrangian points that leads to introduce orbital perturbations represented by $\zeta$  and adds a new frequency in the system, and (iii) the combination $\sqrt{\mu} \rho$ coming from the short-period orbital perturbation.

(i) The coorbital satellites are small satellites of some kilometers (the mean radius ranges from  1.3 km for Polydeuces to 89.5 km for Janus). The Cassini spacecraft provides high resolution images leading to mapping and cartography these satellites with an accuracy of 0.4 km for Polydeuces \citep{PoThoWeRi2007, Thomas2010}. The only accessible parameters for the figure of the satellites are the long axes ellipsoidal shape. From these data, we can compute the moment of inertia by assuming that the satellites are homogeneous and ellipsoidal \begin{eqnarray}
A = \frac{ M}{5} (b^2 + c^2) ,\\
B = \frac{ M}{5} (c^2 + a^2) ,\\
C = \frac{ M}{5} (a^2 + b^2) ,
\end{eqnarray}
where $A,B,C$ are the principal moment of inertia and $a,b,c$ are the fitted long axis ellipsoidal shape coming from shape models \citep{Thomas2010}. By looking the recent images of Helene (Cassini image N00172886.jpg, NASA/JPL/Space Science Institute), it is clear that it is a rough hypothesis. We use the data provided by \cite{Thomas2010} that compiled the recent shape models of these satellites and we compute the triaxiality $(B-A)/C$ that is equal to 
\begin{equation}
\frac{B-A}{C} = \frac{a^2 - b^2}{a^2 + b^2},
\end{equation}
so the main uncertainties come from the equatorial axes of the shape. The resulting values of $\sigma = n \sqrt{{3(B-A)/C}}$ are shown in Table~\ref{tab:parameters}. To give an idea of the error due to the ellipsoidal approximation, it is interesting to review the determination of the shape of Janus and Epimetheus. \cite{TiThBu2009} provided the moments of inertia of both satellites from shape model without the assumption of the ellipsoidal shape and deduced $\sigma$ equal to 8.51, and 4.94 rad/day instead of 5.62, and 3.85 rad/day for the ellipsoidal shape model. Clearly, the ellipsoidal shape model gives an error of the order of 33 \%. The Table~\ref{tab:parameters} provides the different values of $\sigma$ for the coorbitals by assuming an ellipsoidal shape. 
We note that Telesto, Calypso, and Polydeuces present a departure from the 1:1 secondary resonance of 3.3, 12.0, 23.2 $\%$, whereas Helene shows 19.0\% departure from the 1:2 secondary resonance. Consequently, by assuming an uncertainty of 30\%, these satellites could be in or close to secondary resonance and more data are necessary to determine their figure. 
Basing on the values given in the tabels \ref{tab:orb} and \ref{tab:parameters}, we estimate the amplitude of the forced orbit for each of the six coorbital satellites. To this end, we assume that the rotation of the satellites is not in secondary resonance, and use the expression (\ref{eq:sol_force}) to derive the amplitude of the libration with respect to the line Saturn-satellite $A_S = 2\rho\smu n^2(n^2-\sigma^2)^{-1}$, which is given in the fifth column of the table \ref{tab:parameters}. The amplitude of the forced orbit in an uniformly  rotating reference frame (Tab. \ref{tab:parameters}, last column) derived from the expression (\ref{eq:sol_force_rot}) is given by   $A_r = 2\rho\smu \sigma^2(n^2-\sigma^2)^{-1}$.

(ii) The departure of the trajectories from the Lagrange points leads to long-period oscillations in the orbital motion characterized by $\zeta$. 
The perturbation of the orbital motion influences the rotational motion of the satellites by varying the gravitational torque coming from Saturn expressed in (\ref{eq:sol_orb_tau}), through $r$ and $v$. This is an indirect effect acting on the satellite.
It has been shown in the equation (\ref{eq:sol_force}) that the amplitude of the forced trajectory is almost the same than in the Keplerian case but its phase differs by the $\nu$-periodic function $\zeta(\nu t)$. For the $\sigma$ values listed in the Table~\ref{tab:parameters}, the influence of $\zeta$ is small for Calypso, Telesto, and Helene. However, for large tadpole orbits such as Polydeuces and for horseshoe orbits (Janus and Epimetheus) the influence is large.
In addition, for $\sigma$ leading to secondary spin-orbit resonance, the presence of the additional frequency $\nu$ causes new resonances and generates chaotic areas.  

(iii) The orbital motion of the satellites is described through the \'Erdi's formalism (\'Erdi 1977) and the initial values come from the \textit{Horizons} ephemerides \citep{Gioetal1996}. The \textit{Horizons} ephemerides provide a recent orbit of the satellites based on additional images from Cassini spacecraft \citep{Jacobsonetal2008}. In our model the parameter $\sqrt{\mu} \rho$ controls the amplitude of the orbital short period variations  as the eccentricity does for Keplerian motions. Consequently, $\rho$ is directly adjusted to the mean eccentricity of the satellite. The eccentricity of the coorbitals from the \textit{Horizons} ephemerides varies with time on short and long periods due to several effects. First, the eccentricity of the principal satellites is not equal to zero. The eccentricities of Dione and Tethys are 0.0022 and 0.0001, respectively. \'Erdi (1977) shows that in this case, the Equation~(2) will present an additional periodic term at the orbital frequency leading to an additional term of frequency $n$ in the solution. Moreover, due to the interactions between the satellites, the longitude $v$ will present secular terms, long periodic component, as illustrated for Janus and Epimetheus \citep{RoRaCa2011}. All these effects could be easily introduced in the present formalism by adding periodic terms in the equation of motion. On contrast, the interactions with Enceladus and Mimas that are in 2:1 resonance with Dione and Tethys, and therefore with the coorbitals, are more complicated to model and to describe. 

\begin{table}[h]
\begin{center}
\begin{tabular}{lccccc}
\hline
\text{Satellite} & $n$\phantom{xxx} & $\sigma$& $\smu\rho$ & $A_S$ & $A_r$ \\ 
 & (rad/day)& (rad/day) & & (deg.) & (deg.) \\ 
\hline
Polydeuces &$2.3$ &$ 1.86$ &0.0196&$6.5$ & $4.2$ \\
Helene         & $2.3$ &$1.42$ &0.0078&$1.5$ & $0.6$  \\
Telesto         & $3.3$ &$3.22$ &0.0006 &$1.4$ & $1.4$  \\
Calypso       & $3.3$ &$2.97 $ &0.0008&$0.5$ & $0.4$  \\
Janus$^{(a)}$           &  $9.03$&$4.94$ &0.0068&$1.1$ & $0.3$  \\
Epimetheus$^{(a)}$ & $9.03$&$8.51$ &0.0098 & $10$ &  $8.9$  \\
\hline
\end{tabular}
\end{center}
\caption{Amplitude of the forced orbit. The mean motion $n$ and the parameter $\sigma$ (libration frequency at the center of libration) are given in the second and third columns.  The fourth column corresponds to the values of the parameter $\smu\rho$ that governs the variations of the orbital motion at the frequency $n$. $A_S$ is the amplitude of the forced orbit measured with respect to the line Saturn-satellite, while $A_r$ represent the amplitude of the short-period term of the forced trajectory in uniformly rotating frame. The symbol  ${(a)}$ indicates that  the value of $\sigma$ comes from \cite{TiThBu2009}. 
}
\label{tab:parameters}
\end{table}%

In conclusion, we have studied the impact of the coorbital resonance on the rotational motion of Telesto, Calypso, Helene and Polydeuces, in the planar problem (orbital and equatorial inclination are neglected). We have shown that the main parameters acting on the rotational motion are ($\sigma,\zeta,  \sqrt{\mu}\rho $) where $\sigma$ is related to the dynamical figure of the moons, $\zeta$ is the secular perturbation of the orbit and $\sqrt{\mu}\rho$ is the departure from the Lagrangian points. The uncertainties in the dynamical figure of the moons range the possibility that they are close or in secondary resonances. Such secondary resonances, coupled with the secular motion of the moons due to the coorbital resonance, increase the possibility to reach supplementary resonances and these satellites are inclined to exhibit a rich dynamics. The determination of the dynamical figure of these satellites by space observations will bring precious information for the dynamics of these moons.

\section*{Appendix}
The link with the expressions given by \cite{Ed1977} and the  equations (\ref{eq:sol_orb_r}) to (\ref{eq:sol_orb_zeta}) is not straightforward and some clues are given in the following lines. The $\text{x}^{\text{th}}$ formula is denoted by [x] if it appears in \cite{Ed1977} and by (x) if it comes from the present paper.

From the relations [5a] and [19] truncated at  first order in $\varepsilon = \smu$, we deduce that $r = 1 +\smu\left(\rho_1\cos(\varphi + \psi_1) -\frac23 f_1\right)$, while [5b] and [20] give 
$\theta = \varphi -  2\smu\rho_1\sin(\varphi +\psi_1) + \smu q_1$ (the polar angle $\theta$ is denoted by $v$ in our paper). The functions $\varphi, f_1, \rho_1$ and $\psi_1$ depend a priori on $\tilde v$ which is a slow time defined by $\tilde v = \smu(v - v_0)$, $v$ being the true anomaly of the massless body ($v = nt$ in the Circular Restricted Three Body Problem (C.R.T.B.P.)).
As from [27a] and [27b], the functions $\rho_1$ and $\psi_1$ are constant ($e_1=0$ in the C.R.T.B.P.),  it only remains to give an explicit expression for $\varphi$ and $f_1$ .
From [7a] we have $\varphi = v + \smu\int f_1(\tilde v) dv$ and the relation [23] links $f_1$, and consequently  $\varphi$,  to $\zeta_1$ ($\zeta$ in the present paper) that satisfies the differential equation [24] (equation (5) in this text). By some changing of notations, we get the approximations  (\ref{eq:sol_orb_r}) to (\ref{eq:sol_orb_zeta}).

Here, we have extracted from \'Erdi's paper only the necessary relations to get a simple orbital model of a Trojan satellite. However,  \'Erdi's theory goes far further since it gives asymptotical  expansions to second order in $\smu$ within the framework of the planar elliptic restricted three body problem. The section 7 of \cite{Ed1977} summarizes these results. A spatial theory is given in \cite{Er1978}.

\begin{acknowledgements}
The authors thank Elodie Thilliez for her help concerning the  application of \'Erdi's approximations and  the referee Benoit Noyelles  for his useful remarks and suggestions.\end{acknowledgements}


\newcommand{\noopsort}[1]{}

\end{document}